\documentclass[12pt, draftclsnofoot,onecolumn]{IEEEtran}

\usepackage{xspace,amsmath,amssymb,amsfonts,epsfig,syntonly}
\usepackage{cite,bm,color,url,textcomp,empheq,boxedminipage}
\usepackage{algorithmicx,algorithm,algpseudocode}
\usepackage{epstopdf}
\usepackage{empheq}
\usepackage{graphicx,graphics}  
\usepackage{multirow,multicol}
\usepackage{psfrag}    
\usepackage{stfloats}
\usepackage{url}

\newtheorem{lemma}{\underline{Lemma}}[section]

\newtheorem{proposition}{\underline{Proposition}}[section]

\newtheorem{remark}{\underline{Remark}}[section]

\newcommand{\mv}[1]{\mbox{\boldmath{$ #1 $}}}

\DeclareMathOperator*{\argmin}{arg\,min}

\long\def\symbolfootnote[#1]#2{\begingroup
\def\thefootnote{\fnsymbol{footnote}}
\footnote[#1]{#2}\endgroup}

\psfull

\usepackage[T1]{fontenc}
\usepackage{ifpdf}
 \ifpdf
 \else
 \fi

\begin{document}
\title{Multi-Antenna NOMA for Computation Offloading in Multiuser Mobile Edge Computing Systems}

\author{Feng Wang, Jie Xu, and Zhiguo Ding
\thanks{Part of this paper has been presented at the IEEE Global Communications Conference Workshop on Non-Orthogonal Multiple Access Techniques for 5G, Singapore, December 4--8, 2017\cite{conf_version}.}

\thanks{F. Wang and J. Xu are with the School of Information Engineering, Guangdong University of Technology, Guangzhou 510006, China (e-mail: \{fengwang13, jiexu\}@gdut.edu.cn). J. Xu is the corresponding author.}
\thanks{Z. Ding is with the School of Electrical and Electronic Engineering, the University of Manchester, Manchester, M13 9PL, UK (e-mail: zhiguo.ding@manchester.ac.uk).}}

\maketitle

\begin{abstract}
This paper studies a multiuser mobile edge computing (MEC) system, in which one base station (BS) serves multiple users with intensive computation tasks. We exploit the multi-antenna non-orthogonal multiple access (NOMA) technique for multiuser computation offloading, such that different users can simultaneously offload their computation tasks to the multi-antenna BS over the same time/frequency resources, and the BS can employ successive interference cancellation (SIC) to efficiently decode all users' offloaded tasks for remote execution. In particular, we pursue energy-efficient MEC designs by considering two cases with partial and binary offloading, respectively. We aim to minimize the weighted sum-energy consumption at all users subject to their computation latency constraints, by jointly optimizing the communication and computation resource allocation as well as the BS's decoding order for SIC. For the case with partial offloading, the weighted sum-energy minimization is a convex optimization problem, for which an efficient algorithm based on the Lagrange duality method is presented to obtain the globally optimal solution. For the case with binary offloading, the weighted sum-energy minimization corresponds to a {\em mixed Boolean convex problem} that is generally more difficult to be solved. We first use the branch-and-bound (BnB) method to obtain the globally optimal solution, and then develop two low-complexity algorithms based on the greedy method and the convex relaxation, respectively, to find suboptimal solutions with high quality in practice. Via numerical results, it is shown that the proposed NOMA-based computation offloading design significantly improves the energy efficiency of the multiuser MEC system as compared to other benchmark schemes. It is also shown that for the case with binary offloading, the proposed greedy method performs close to the optimal BnB based solution, and the convex relaxation based solution achieves a suboptimal performance but with lower implementation complexity.
\end{abstract}


\section{Introduction}
Recent advancements in smart Internet of things (IoT) have motivated various computation-intensive and latency-critical applications such as virtual reality, augmented reality, autonomous driving, unmanned aerial vehicles (UAVs), and tele-surgery \cite{Chiang16}. The success of these applications requires future wireless networks to incorporate billions of IoT devices for real-time communication, computation, and control. However, as IoT devices generally have small physical sizes and limited power supply, it is a challenging task for them to handle intensive computation loads with critical latency requirements. Conventionally, the mobile cloud computing (MCC) technique is employed to provide IoT devices with abundant computation resources at (remote) centralized cloud. However, due to the long distances between the IoT devices and the centralized cloud, MCC may result in significantly long computation latency and incur heavy traffic loads at the backhaul networks.

To resolve this issue, recently mobile edge computing (MEC) has been proposed as an alternative solution\cite{Chiang16, Bar14,Mach17}, which provides cloud-like computing at the network edge (e.g., cellular base stations (BSs)) by deploying edge servers therein. With MEC, these IoT devices can perform {\em computation offloading} to transfer their computation-intensive tasks to the BS; after successful remote execution by the edge server therein, the BS then sends the computation results back to the devices\cite{Mach17}. As the BS is located in close proximity to the IoT devices, MEC is able to considerably shorten their computation latency, and significantly reduce the traffic loads at the backhaul networks. As such, MEC has attracted extensive attentions from both academia and industry in the smart IoT era\cite{Mao17,ETSI}.

The practical implementation of MEC, however, faces various technical challenges. First, computation tasks at users can be classified into different categories depending on the tasks' partitionability and dependence. As a result, the computation offloading must be designed based on the corresponding task models\cite{Mao17}. For example, partial offloading and binary offloading are two widely adopted computation offloading models in the MEC literature\cite{Mao17}, in which the tasks at each user are fully partitionable and non-partitionable, respectively. Next, the performance optimization of computation offloading in MEC systems critically relies on the joint design of both communication and computation resource allocations\cite{Wang16,Simeone17,Munoz15,Huang12,Liu16}. For example, consider that one BS serves one single actively-computing user. In order to minimize the energy consumption for task execution, it is crucial for the user to jointly optimize the communication power (for offloading) and the central processing unit (CPU) frequencies for local computing to balance their energy consumption tradeoff. For the case of partial offloading, the user needs to properly partition the computation task into two parts for offloading and local computing, respectively; for the case of binary offloading, the user needs to properly choose the operation mode between offloading and local computing for energy minimization. Furthermore, future wireless networks are expected to consist of massive IoT devices, and each BS generally needs to serve a large number of IoT devices at the same time for executing their offloaded computation tasks. How to efficiently and fairly share the wireless communication resources (for offloading) and the edge server's computation resources (for remote execution) among multiple users in the joint communication and computation optimization is a highly challenging task to be tackled.

In the literature, there have been various prior works studying the joint communication and computation design in multiuser MEC systems\cite{Wang17, Chen18,Bi17,Mao17_2,Sar15,ChenXu16,You17,Feng17,Cao17,Hu18}. For example, the authors in \cite{Chen18,Mao17_2,Sar15} and \cite{You17} studied the orthogonal frequency-division multiple access (OFDMA) based multiuser computation offloading for the cases with binary and partial offloading, respectively, in which the communication and computation resource allocations are jointly optimized to minimize the users' sum-energy under different setups. \cite{Wang17} considered the OFDMA based multiuser computation offloading jointly with the caching technique to maximize the system utility. \cite{ChenXu16} utilized the game theory to investigate the energy efficiency tradeoff among different users in a multiuser MEC system with the code-division multiple access (CDMA) based offloading. \cite{Feng17,Bi17} considered a wireless powered MEC system with time-division multiple access (TDMA) based offloading, in which the computation offloading and local computing at the users are powered by wireless power transfer (WPT) from the BS. Furthermore, \cite{Cao17,Hu18} studied a new communication and computation cooperation approach in an MEC system consisting of one user, one helper, and one BS, in which a TDMA based offloading protocol is proposed, such that the user can explore the communication and computation resources at both the helper and the BS for computation performance optimization. However, despite the research progress, the above prior works only considered generally suboptimal multiuser computation offloading schemes by using orthogonal multiple access (OMA) for computation offloading (e.g., TDMA and OFDMA) or employing CDMA by treating interference as noise. These schemes, however, cannot fully explore the capacity of the multiple access channel for offloading from multiple users to the BS, and thus may lead to suboptimal performance for multiuser MEC systems. This thus motivates us to investigate new multiple access schemes for multiuser offloading in this paper.

Recently, non-orthogonal multiple access (NOMA) has been recognized as one of the key techniques in the fifth-generation (5G) cellular networks \cite{Ding2016,Ding16,Ding17,Dai15,Ng17}. Unlike conventional OMA, NOMA enables multiple users to communicate with the BS at the same time and frequency resources. By implementing sophisticated multiuser detection schemes such as the {\em successive interference cancellation} (SIC) at receivers\cite{TseBook,Rui06,Tse98}, the NOMA-based communication system is expected to achieve a much higher spectral efficiency than the OMA counterpart. In particular, for a single-cell uplink NOMA system or equivalently a multiple access channel from users to the BS, it has been well established that the information-theoretical capacity region is achievable when the users employ Gaussian signaling with optimized coding rates, and the BS receiver adopts the minimum mean square error (MMSE)-SIC decoding with a properly designed decoding order for different users (see, e.g., \cite{TseBook}). Motivated by the benefit of NOMA over OMA, it is expected that NOMA can be exploited to further improve the performance of multiuser computation offloading for MEC systems.

In this paper, we investigate the NOMA-based multiuser computation offloading designs for a multiuser MEC system, which consists of one multi-antenna BS (integrated with an edge server) and multiple single-antenna users. Each user has certain computation tasks that need be successfully executed within a particular finite-duration block. Based on the uplink NOMA protocol for computation offloading, these users can simultaneously offload their tasks to the BS over the same time/frequency resources. In particular, we pursue an energy-efficient MEC design by considering two scenarios with partial and binary offloading, respectively. For both cases, we aim to minimize the weighted sum-energy at all users while ensuring the successful execution at each user within this block, by jointly optimizing the users' offloading decision, CPU frequencies for local computing, transmission powers and rates for offloading, as well as the BS's decoding order for MMSE-SIC.

For the case with partial offloading, the weighted sum-energy minimization is formulated as a convex optimization problem. Nonetheless, this problem does not admit explicit functions for the users' offloading rates due to the initially unknown decoding order at the BS; and therefore, it is generally difficult to find the optimal solution efficiently. By applying the Lagrange duality method and leveraging the {\em polymatroid} structure of the capacity region of the multiple access channel\cite{TseBook}, we present an efficient algorithm to obtain the globally optimal solution to this problem.

For the case with binary offloading, the weighted sum-energy minimization corresponds to a mixed Boolean convex problem that is generally NP-hard \cite{Wig06} and thus more difficult to be solved. First, we propose a branch-and-bound (BnB) based algorithm to obtain the globally optimal solution, which, however, has very high implementation complexity especially when the number of users becomes large. Next, we propose two low-complexity algorithms based on the greedy method and the convex relaxation, respectively, to obtain suboptimal solutions with high quality in practice.

Finally, we present numerical results to validate the performance of our proposed multi-antenna NOMA-based computation offloading designs. For both cases with partial and binary offloading, it is shown that the proposed NOMA-based offloading design achieves substantial energy efficiency gains as compared to the benchmark schemes with OMA-based offloading, local computing only, and full offloading only. It is also shown that for the case with binary offloading, the proposed greedy method performs close to the optimal BnB based solution, while the convex relaxation based method achieves a suboptimal performance but with much lower implementation complexity.

The remainder of this paper is organized as follows. Section~\ref{sec:system} presents the multiuser MEC system model with multi-antenna NOMA-based computation offloading, and formulates the weighted sum-energy minimization problems for the cases with partial and binary offloading, respectively. Section~\ref{sec:optimal} proposes an efficient algorithm to obtain the globally optimal solution to the problem for the case with partial offloading. Section~\ref{sec:binary} presents both optimal and suboptimal solutions to the problem for the case with binary offloading. Section~\ref{sec:numerical} provides numerical results to evaluate the performance of the proposed NOMA-based offloading designs as compared to other benchmark schemes. Finally, Section~\ref{sec:conclusion} concludes this paper.

{\em Notations:} $\mathbb{C}^{x\times y}$, ${\mathbb R}^{x\times y}$, and  ${\mathbb R}^{x\times y}_+$ denote the sets of all complex-valued, real-valued, and nonnegative real-valued matrices with dimension $x\times y$, respectively; $\bm I$ denotes an identity matrix with appropriate dimension; ${\mathbb E}[\cdot]$ denotes the statistical expectation. $\|\bm x\|$ and $\bm x^\dagger$ denote the Euclidean norm and the transpose of a vector $\bm x$, respectively; $|x|$ denotes the absolute value of a scalar $x$; $|{\cal S}|$ denotes the cardinality of a set $\cal S$; $\det(\bm A)$ denotes the determinant of a square matrix $\bm A$. The distribution of a circularly symmetric complex Gaussian (CSCG) random vector $\bm x$ with mean $\bm \mu$ and covariance $\bm \Sigma$ is denoted by $\bm x\sim {\cal CN}(\bm \mu,\bm \Sigma)$, where $\sim$ stands for ``distributed as''. Furthermore, we define $(x)^+=\max(x,0)$.

\section{System Model}\label{sec:system}


\begin{figure}
  \centering
  \includegraphics[width = 3.5in]{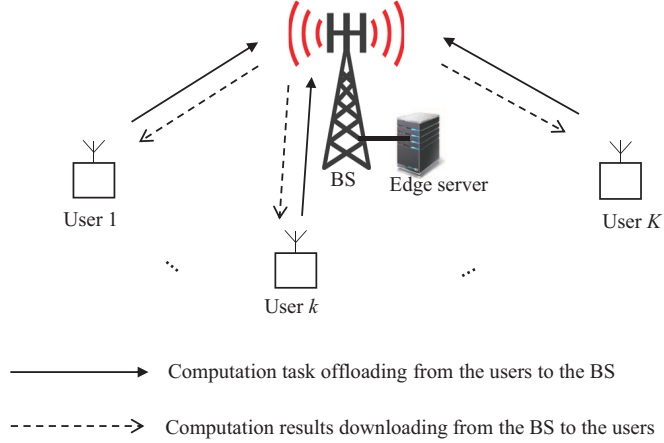}
    \caption{System model for the multiuser MEC system.} \label{fig.SystemMod}
\end{figure}
\begin{figure}
  \centering
  \includegraphics[width =3.5in]{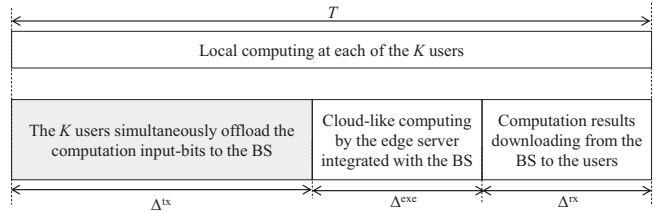}
    \caption{NOMA-based computation offloading protocol in the multiuser MEC system.} \label{fig.TBprotocol}
\end{figure}

We consider an MEC system consisting of one single BS equipped with $N\geq1$ antennas and a set ${\cal K}\triangleq \{1,\ldots,K\}$ of users each with a single antenna, as shown in Fig.~\ref{fig.SystemMod}. The BS is integrated with an edge server that can execute the offloaded computation tasks from the $K$ users. We focus on a particular block with finite duration $T$, and each user $k\in{\cal K}$ has a computation task with $L_k$ input-bits in total that should be executed before the end of this block. Here, $T$ is chosen to be no larger than the channel coherence time, such that the wireless channels remain unchanged for all the users during this block. We consider two cases with partial and binary offloading. For partial offloading, the task at each user $k\in{\cal K}$ can be arbitrarily partitioned into two parts for offloading and local computing, respectively. For binary offloading, the task at user $k\in{\cal K}$ can only be executed as a whole via either offloading or local computing. Without loss of generality, at user $k\in{\cal K}$, we denote $\ell_k$ as the number of task input-bits for offloading, and denote $(L_k-\ell_k)$ as the number of task input-bits for local computing. Therefore, for partial offloading, $\ell_k$ can be viewed as a continuous variable between 0 and $L_k$ for any $k\in{\cal K}$.\footnote{Note that the number of task input-bits $\ell_k$ generally should be an integer variable. Nevertheless, it is a safe approximation to view $\ell_k$ as a continuous variable for the case with partial offloading where each task input-bit can be viewed as an independent task, especially when the number $L_k$ of task input-bits is large.} For binary offloading, we have $\ell_k\in\{0,L_k\}$, $k\in{\cal K}$. Note that in this paper, we assume that the BS collects both the computation information of all the $K$ users and the channel state information (CSI) from/to the $K$ users, and accordingly, the BS can coordinate the computation offloading and local computing for these users.

\vspace{-0.3cm}
\subsection{NOMA-based Multiuser Computation Offloading}
The NOMA-based multiuser computation offloading protocol is implemented as shown in Fig.~\ref{fig.TBprotocol}, where the duration-$T$ block of our interest is divided into three phases, for the users' task offloading in the uplink, the BS's remote task execution, and the uses' computation results downloading in the downlink, respectively. In the first phase with duration $\Delta^{\text{tx}}$, all the $K$ users simultaneously offload their computation tasks, with $\ell_k$ input-bits for each user $k\in{\cal K}$, to the BS based on NOMA. After collecting these bits, in the second phase with duration $\Delta^{\rm exe}$, the edge server at the BS remotely executes the offloaded tasks on behalf of these users by developing multiple virtual machines (VMs) each for one user\cite{Mach17}. In the third phase with duration $\Delta^{\rm rx}$, the BS sends the computation results back to these $K$ users. In order for each user $k\in{\cal K}$ to obtain the computation results before the end of this block, we have\footnote{Note that in order for each user $k\in{\cal K}$ to transmit a given number $\ell_k$ of task input-bits to the BS, the user's transmission energy consumption is monotonically decreasing with respect to the transmission duration\cite{Gamal02}. On the other hand, the user $k$'s local computing energy consumption for executing a given number $(L_k-\ell_k)$ of task input-bits is monotonically decreasing with respect to the task execution time\cite{Bar14}, as shown in (10) later. By combining the two facts, the strict equality in \eqref{eq.Delta} must hold in order to minimize the users' weighted energy consumption while ensuring their successful tasks execution.}
\begin{align}\label{eq.Delta}
\Delta^{\rm tx} + \Delta^{\rm exe} + \Delta^{\rm rx} = T,
\end{align}
where the strict equality is set for minimizing the related communication and computation energy consumption \cite{Bar14,Gamal02}. As the edge server generally has sufficient computation resources, it can adopt a significantly high CPU frequency to minimize the execution time for the offloaded tasks from the users. Furthermore, the computation results are generally of smaller size than the computation task input-bits (e.g., speech recognition, image rendering, and feature extraction in the augmented reality based applications) and the BS can usually use high transmit power to send the computation results to the intended users. For the purpose of exposition, we focus our study on the computation offloading phase, by assuming both the time duration $\Delta^{\rm exe}$ for the execution and the duration $\Delta^{\rm rx}$ for computation results downloading to be constant. Therefore, it follows from \eqref{eq.Delta} that the transmission time for offloading is given as a constant as
\begin{align}\label{eq.Delta_2}
\tilde{T} \triangleq \Delta^{\rm tx} = T -\Delta^{\rm exe} - \Delta^{\rm rx}.
\end{align}

Now, we focus on the multiuser computation offloading in the first phase. The $K$ users employ uplink NOMA to offload their respective task input-bits to the BS simultaneously. Let $x_k\in{\mathbb C}$ denote the user $k$'s task-bearing signal with unit power for offloading, i.e., $\mathbb{E}[|x_k|^2]=1$, and $p_k>0$ denote the corresponding transmit power, $k\in{\mathcal K}$. The received signal $\bm y\in\mathbb{C}^{N\times 1}$ at the BS is then expressed as \cite{Ding16}
\begin{align}\label{eq.mac}
\boldsymbol{y} = \sum_{k=1}^{K} \sqrt{p_k} \boldsymbol{h}_k x_k+\boldsymbol{z},
\end{align}
where $\boldsymbol{h}_k\in{\mathbb C}^{N\times 1}$ denotes the uplink channel vector from user $k$ to the BS and $\boldsymbol{z}\in{\mathbb C}^{N\times 1}$ denotes the additive white Gaussian noise at the BS, which is normalized to be of unit power with $\boldsymbol{z}\sim{\cal CN}(0,\boldsymbol{I})$.

At the receiver side, the BS employs the MMSE-SIC to decode information from the $K$ users. Let the permutation $\bm \pi$ over $\cal K$ denote the {\em successive decoding order} at the BS, which indicates that the BS receiver first decodes the information $x_{\pi(K)}$ transmitted by user $\pi(K)$, then decodes $x_{\pi(K-1)}$ by cancelling the interference from $x_{\pi(K)}$, followed by $x_{\pi(K-2)}$, $x_{\pi(K-3)}$, and so on, until $x_{\pi(1)}$. By employing the capacity-achieving Gaussian signaling at these users (i.e., setting $x_k$'s to be independent CSCG random variables), the achievable rate (in bits/sec) at user $\pi(k)$ under a given decoding order $\bm \pi$ and a set of the users' transmit powers powers $\{p_k\}$ is given by\cite{TseBook,Tse98,Rui06}
\begin{align}\label{eqn:r:pi:k}
r_{\pi(k)}^{(\pi)} = B\log_2\left(\frac{\det\left(\bm I + \sum_{i=1}^k p_{\pi(i)} \bm h_{\pi(i)} \bm h^H_{\pi(i)} \right)}{\det\left( \bm I + \sum_{i=1}^{k-1}p_{\pi(i)} \bm h_{\pi(i)}\bm h_{\pi(i)}^H\right)}\right), \quad k\in{\cal K},
\end{align}
where $B$ denotes the transmission bandwidth for the users. By allowing the BS to properly design the decoding order and employ time-sharing among different decoding orders, the capacity region for the $K$ users is achievable and corresponds to the following polymatroid \cite{TseBook}:
\begin{align}\label{eq.C}
{\cal X}(\bm p)& \triangleq \left\{ \bm r\in{\mathbb R}_+^{K\times1}: \sum_{k\in{\cal J}} r_k \leq B\log_2\left( \det\left( \bm I +\sum_{k\in{\cal J}}p_k\bm h_k\bm h_k^H\right)\right),\;\forall {\cal J}\subseteq{\cal K}
\right\},
\end{align}
where $r_k\geq 0$ denotes the achievable rate of user $k\in{\cal K}$, $\bm r \triangleq [r_1,\ldots,r_K]^{\dagger}$, $\mv p\triangleq[p_1,\ldots,p_K]^\dagger$, and ${\cal J}$ denotes any subset contained in set $\cal K$. 

Over the block with task offloading duration $\tilde{T}$, the maximum number of bits that can be offloaded from user $k$ to the BS is given by $r_k\tilde{T}$. In order for user $k\in{\cal K}$ to successfully offload the tasks to the BS, $r_k\tilde{T}$ should be no smaller than the number $\ell_k$ of offloaded task input-bits. Therefore, we have
\begin{align}\label{eq.rate_cons}
r_k \tilde{T} \geq \ell_k,~~ k\in{\cal K}.
\end{align}
Furthermore, we consider the transmission energy consumption as the sole energy budget at user $k\in{\cal K}$ for computation offloading, which is expressed as
\begin{align}\label{eq.Ek-tx}
E_k^{\rm tx} = p_k\tilde{T}.
\end{align}

\vspace{-0.4cm}
\subsection{Local Computing at Users}
Next, we consider the local computing of the $(L_k-\ell_k)$ task input-bits at each user $k\in{\cal K}$. In general, the number of CPU cycles required for executing a task depends on various issues such as the number of its input-bits, the specific application, as well as the CPU and memory architectures at the user\cite{Burd96}. For ease of exposition and as commonly adopted in the MEC literature (see, e.g., \cite{Liu16,Wang17, Chen18,Mao17_2,Sar15,ChenXu16,You17,Feng17}), we assume that the number of CPU cycles is linear with the number of task input-bits. Let $C_k$ denote the number of CPU cycles required for computing one input-bit at each user $k\in\mathcal K$. Accordingly, a total number of $C_k(L_k-\ell_k)$ CPU cycles is required for user $k$ to compute $(L_k-\ell_k)$ task input-bits.

In order to minimize the energy consumption for local computing, user $k\in{\cal K}$ applies the dynamic voltage and frequency scaling (DVFS) technique\cite{Mach17} to adaptively adjust the CPU frequency $f_{k,q}$ for each CPU cycle $q\in\{1,\ldots,C_k(L_k-\ell_k)\}$. As a result, the execution time for one CPU cycle $q$ at user $k$ is given as $1/f_{k,q}$. In this case, to ensure local computing to be accomplished before the end of this block, the CPU frequencies $\{f_{k,q}\}$ should satisfy the following computation latency constraints:
\begin{align}\label{eq.f_in}
\sum_{q=1}^{C_k(L_k-\ell_k)} \frac{1}{f_{k,q}}\leq T,~~\forall k\in {\cal K}.
\end{align}
Consequently, the energy consumption for local computing at user $k\in{\cal K}$ is given by\cite{Mach17}
\begin{align}\label{eq.energy_loc}
E^{\rm loc}_k =\sum_{q=1}^{C_k(L_k-\ell_k)} \zeta_k f^2_{k,q},
\end{align}
where $\zeta_k>0$ is the effective capacitance coefficient that depends on the chip architecture at user $k$.

Note that the computation delay in the left-hand-side (LHS) of \eqref{eq.f_in} and the energy consumption in the right-hand-side (RHS) of \eqref{eq.energy_loc} are both convex with respect to the CPU frequencies $\{f_{k,q}\}$. As a result, $\{f_{k,q}\}$ should be set to be identical for different CPU cycles $q\in\{1,\ldots,C_k(L_k-\ell_k)\}$ to minimize the energy consumption while ensuring the computation latency (see \cite[Lemma 3.1]{Feng17}). By using this fact and setting the computation latency constraints in \eqref{eq.f_in} to be met with strict equality, we have $f_{k,1}=\ldots=f_{k,C_k(L_k-\ell_k)}=C_k(L_k-\ell_k)/T$ for any $k\in{\cal K}$. Substituting this in \eqref{eq.energy_loc}, the energy consumption $E^{\rm loc}_k$ for local computing at user $k\in{\cal K}$ is re-expressed as
\begin{align}\label{eq.Ek-loc}
E^{\rm loc}_k =\frac{\zeta_kC_k^3(L_k-\ell_k)^3}{T^2}.
\end{align}

\vspace{-0.4cm}
\subsection{Problem Formulation}\label{sec:problem}
To pursue an energy-efficient MEC design, we are interested in minimizing the weighted sum-energy consumption at the $K$ users, i.e., $\sum_{k=1}^K \alpha_k(E^{\rm tx}_k+E^{\rm loc}_{k})$ with $E^{\rm tx}_k$ in \eqref{eq.Ek-tx} and $E^{\rm loc}_{k}$ in \eqref{eq.Ek-loc}, while ensuring that the computation tasks at the $K$ users are successfully executed within this block. Here, $\mv \alpha\triangleq[\alpha_1,\ldots,\alpha_K]^\dagger\in{\mathbb R}_+^{K\times1}$ denotes the given {\em energy-weight} vector for characterizing the priority of different users in energy minimization. The decision variables include the $K$ users' task partitions $\mv \ell \triangleq [\ell_1,\ldots,\ell_K]^\dagger$, transmission powers $\mv p$ and rates $\mv r$ for offloading, as well as the associated decoding order $\bm \pi$ at the BS receiver.

\subsubsection{Partial Offloading}
First, we consider the case with partial offloading, in which the number $\ell_k$ of each user $k$'s offloaded task input-bits is a continuous variable within the interval $[0,L_k]$. In this case, the weighted sum-energy minimization problem is formulated as
\begin{subequations}\label{eq.prob1}
\begin{align}
\text{(P1)}:~
 \min_{\mv \ell, \;\mv p,\; \mv r } &~\sum_{k=1}^K \alpha_k\left(\frac{\zeta_kC_k^3(L_k-\ell_k)^3}{T^2}+p_k\tilde{T}\right) \\
{\rm s.t.}~~& {\bm r} \in{\cal X}(\mv p)\\
& r_k \geq \ell_k/\tilde{T},~~\forall k\in{\cal K}\\
& p_k \geq 0,~~\forall k\in{\cal K}\\
& 0\leq \ell_k \leq L_k,~~\forall k\in{\cal K},
\end{align}
\end{subequations}
where the constraints in (\ref{eq.prob1}b) specify that the offloading rate-tuple of the $K$ users must lie within the capacity region of the multiple access channel, under given transmit powers $\mv p$. Notice that for each inequality in ${\cal X}(\mv p)$ (see \eqref{eq.C}), the RHS is concave with respect to $\bm p$ and the LHS is affine with respect to $\bm r$. Therefore, the set of $\bm r$ and $\bm p$ characterized by $\bm r\in{\cal X}(\bm p)$ in (\ref{eq.prob1}b) is convex\cite{Rui06}. Furthermore, the objective function is jointly convex with respect to the nonnegative variables $p_k$'s and $\ell_k$'s and the constraints in (\ref{eq.prob1}c) and (\ref{eq.prob1}d) are all linear inequations. Therefore, problem (P1) is a convex optimization problem. However, problem (P1) is still difficult to be optimally solved, since that problem (P1) does not admit explicit functions for $r_k$'s due to the initially unknown decoding order at the BS receiver. How to infer the optimal decoding order from the optimal solution to problem (P1) is an important but challenging task for practical implementation. Furthermore, based on ${\cal X}(\mv p)$ in \eqref{eq.C}, constraint (\ref{eq.prob1}b) corresponds to a total number of $(2^K-1)$ inequality constraints, which increases exponentially with respect to the user number $K$. Therefore, it is practically infeasible to directly consider these constraints in problem (P1), especially when $K$ becomes large.

\subsubsection{Binary Offloading}
Next, we consider the case with binary offloading, in which $\ell_k\in\{0,L_k\}$ holds for each user $k\in{\cal K}$. In this case, the weighted sum-energy minimization problem is formulated as
\begin{subequations}\label{prob.binary-offloading}
\begin{align}
\text{(P2)}: ~
 \min_{\mv \ell, \mv p,\mv r } &~ \sum_{k=1}^K  \alpha_k\left(\frac{\zeta_kC_k^3(L_k-\ell_k)^3}{T^2}+p_k \tilde{T} \right)\\
{\rm s.t.} ~~& (\ref{eq.prob1}\text{b}),~(\ref{eq.prob1}\text{c}),~\text{and}~(\ref{eq.prob1}\text{d}) \\
& \ell_k\in\{0,L_k\},~~ \forall k\in{\cal K}.
\end{align}
\end{subequations}
For ease of exposition, we define $\xi_k\triangleq \ell_k/L_k$ such that $\xi_k\in\{0,1\}$, $k\in{\cal K}$. By substituting $\ell_k$'s with $\xi_kL_k$ for all $k\in{\cal K}$, problem (P2) is reformulated as the following {\em mixed Boolean convex problem} \cite{BoydSlides_BnB}:
\begin{subequations}\label{prob.binary-offloading2}
\begin{align}
\text{(P2.1)}: ~
\min_{\mv \xi, \mv p,\mv r } &~ \sum_{k=1}^K  \alpha_k\left(\frac{\zeta_kC_k^3L_k^3(1-\xi_k)^3}{T^2}+p_k \tilde{T} \right)\\
{\rm s.t.} ~~& (\ref{eq.prob1}\text{b})~\text{and}~(\ref{eq.prob1}\text{d}) \\
& r_k\tilde{T}/L_k \geq \xi_k,~~ \forall k\in{\cal K}\\
& \xi_k\in\{0,1\},~~ \forall k\in{\cal K},
\end{align}
\end{subequations}
where $\mv \xi\triangleq [\xi_1,\ldots,\xi_K]^\dagger$ collects all the Boolean variables $\xi_k$'s. The reformulated problem (P2.1) is generally more challenging to be solved than problem (P1). Besides the challenges faced in problem (P1), solving problem (P2.1) requires to further deal with the nonconvex Boolean constraints in (\ref{prob.binary-offloading2}d). Indeed, due to the constraints in (\ref{prob.binary-offloading2}d), the mixed Boolean convex problem (P2.1) is an NP-hard problem~\cite{BoydSlides_BnB}, as stated in the following proposition.

\begin{proposition} \label{prop.NP-hard}
Problem (P2.1) (or problem (P2)) is an NP-hard problem.
\end{proposition}

\begin{IEEEproof}
See Appendix A.
\end{IEEEproof}


\section{Optimal Solution to Problem (P1)}\label{sec:optimal}

In this section, we present the optimal solution to problem (P1). As problem (P1) is convex and satisfies the Slater's condition, strong duality holds between problem (P1) and its dual problem. In the following, we leverage the Lagrange duality method to obtain the optimal solution to problem (P1).

Let $\lambda_k \ge 0$, $k\in{\cal K}$, denote the dual variable associated with the $k$th constraint in \eqref{eq.rate_cons}, and define $\bm\lambda\triangleq [\lambda_1,\ldots,\lambda_K]^\dagger$. The partial Lagrangian of problem (P1) is given by
\begin{align}\label{eq.Lag}
{\cal L}(\mv \ell,\mv p,\mv r,\bm \lambda) = &\sum_{k=1}^K \alpha_k\left(\frac{\zeta_kC_k^3(L_k-\ell_k)^3}{T^2}+p_k\tilde{T}\right)+\sum_{k=1}^{K} \lambda_k\left({\ell_k}/{\tilde{T}}-r_k\right) \notag \\
=& \sum_{k=1}^K \tilde{T}\alpha_kp_k -\sum_{k=1}^K \lambda_kr_k  + \sum_{k=1}^K \left(\frac{\alpha_k\zeta_kC_k^3}{T^2}(L_k-\ell_k)^3 +{\lambda_k}\ell_k/\tilde{T}\right).
\end{align}
The dual function is then defined as
\begin{subequations}\label{eq.dual_func}
\begin{align}
g(\mv \lambda) = \min_{\mv \ell,\; \mv p, \; \mv r} &~~{\cal L}(\mv \ell, \mv p,\mv r,\mv \lambda)\\
 {\rm s.t.}~~&\mv r\in{\cal X}(\mv p)\\
 &p_k\geq 0,~~\forall k\in{\cal K}\\
& 0\leq \ell_k\leq L_k,~~\forall k\in{\cal K}.
\end{align}
\end{subequations}
Correspondingly, the dual problem of (P1) is
\begin{subequations}\label{eq.dual_prob}
\begin{align}
{\rm (D1)}: ~ \max_{\mv \lambda}&~~ g(\mv \lambda)\\
 {\rm s.t.}~&~ \lambda_k\geq 0,~~\forall k\in{\cal K}.
\end{align}
\end{subequations}
In the following, we solve problem (P1) by first evaluating $g(\bm \lambda)$ in \eqref{eq.dual_func} under given $\bm \lambda$, and then solving problem (D1) to find the optimal $\bm \lambda$. Denote by $(\mv \ell^{\rm opt},\mv p^{\rm opt}, \mv r^{\rm opt})$ and $\bm\lambda^{\rm opt}$ the optimal solutions to problems (P1) and (D1), respectively.

\subsection{Evaluating $g(\bm\lambda)$ by Solving Problem \eqref{eq.dual_func}}
First, we obtain the dual function $g(\bm \lambda)$ under given $\bm\lambda$ by solving problem \eqref{eq.dual_func}. In this case, problem \eqref{eq.dual_func} can be decomposed into the following $(K+1)$ subproblems, where \eqref{eq.fk} corresponds to the $K$ subproblems each for one user $k\in\mathcal K$.
\begin{align}\label{eq.fk}
 \min_{0\leq \ell_k \leq L_k}~&~ \frac{\alpha_k\zeta_kC_k^3}{T^2}(L_k-\ell_k)^3 +{\lambda_k}\ell_k/\tilde{T},\quad k\in{\cal K}.
\end{align}
\vspace{-0.8cm}
\begin{subequations}\label{eq.g0}
\begin{align}
\min_{\mv r,\;\mv p}~&~ \sum_{k=1}^K \tilde{T}\alpha_kp_k-\sum_{k=1}^K \lambda_kr_k\\
{\rm s.t.}~~& \mv r \in{\cal X}(\mv p)\\
&p_k\geq 0,~~\forall k\in{\cal K}.
\end{align}
\end{subequations}
For convenience of presentation, let $\ell_k^*$, $k\in{\cal K}$, and $(\bm p^*,\bm r^*)$ denote the optimal solutions to the $k$th subproblem in \eqref{eq.fk} and the subproblem in \eqref{eq.g0}, respectively.

For each subproblem $k\in{\cal K}$ in \eqref{eq.fk}, $\ell^*_k$ can be obtained based on the Karush-Kuhn-Tucker (KKT) conditions\cite{BoydBook}.
\begin{lemma}\label{lem_L}
The optimal solution $\ell_k^*$, $k\in{\cal K}$, to problem \eqref{eq.fk} is given by
\begin{align}\label{eq.l_opt}
\ell_k^* = \left( L_k- \sqrt{\frac{T\lambda_k}{3\alpha_k\zeta_kC_k^3}}\right)^{+}.
\end{align}
\end{lemma}

\begin{IEEEproof}
See Appendix B.
\end{IEEEproof}

As for problem \eqref{eq.g0}, we first present the following lemma from \cite{Tse98}.
\begin{lemma}\label{lem.r}
For any given $\mv \lambda$ and $\mv p$, the optimal solution to the following problem
\begin{align}
\max_{\mv r}~~ \sum_{k=1}^K \lambda_kr_k \quad {\rm s.t.}~~\mv r\in{\cal X}(\mv p)
\end{align}
is obtained by a {\em vertex} $\mv r^{(\pi)}\triangleq [r_{\pi(1)}^{(\pi)},\cdots,r_{\pi(K)}^{(\pi)}]^\dagger$ of the polymatroid ${\cal X}(\mv p)$, where $r_{\pi(k)}^{(\pi)}$ is given in \eqref{eqn:r:pi:k} and the permutation $\mv \pi=[\pi(1),\ldots,\pi(K)]^\dagger$ is determined such that $\lambda_{\pi(1)}\geq \cdots\geq \lambda_{\pi(K)}\geq 0$.
\end{lemma}

Based on Lemma~\ref{lem.r} and substituting \eqref{eqn:r:pi:k}, problem \eqref{eq.g0} reduces to optimizing the transmit power vector $\bm p$ as follows.
\begin{subequations}\label{eq.prob_max2}
\begin{align}
\max_{\mv p }~ &~ \sum_{k=1}^K\left(-\tilde{T}\alpha_kp_k+ B(\lambda_{\pi(k)}-\lambda_{\pi(k+1)})\log_2\left(\det \left( \bm I +\sum_{i=1}^kp_{\pi(i)}\bm h_{\pi(i)}\bm h^H_{\pi(i)} \right)\right)\right)\\
 {\rm s.t.}~~&  p_k\geq 0,~~\forall k\in{\cal K},
\end{align}
\end{subequations}
where $\lambda_{\pi(K+1)}\triangleq 0$ is defined for notational convenience. Note that problem \eqref{eq.prob_max2} is convex and the optimal solution $\mv p^*$ to problem \eqref{eq.prob_max2} can thus be efficiently obtained by standard convex solvers, e.g., CVX \cite{CVX}. By substituting $\mv p^*$ into \eqref{eqn:r:pi:k}, the optimal offloading rate tuple $\mv r^*$ is obtained.

\begin{remark}\label{remark:1}
Note that the optimal transmit power $\mv p^*$ for offloading is unique due to the strict convexity of problem \eqref{eq.prob_max2} (also see \cite{Rui06}). However, the offloading rate tuple $\mv r^*$ is generally non-unique. This is because that if there exist any two users $i$ and $j$ such that $\lambda_i = \lambda_j$, $i\neq j$, $i,j\in\mathcal K$, then the two decoding orders (i.e., decoding the message of user $i$ first followed by that of user $j$, and the reverse order) are both optimal for problem \eqref{eq.g0}. Suppose that ${\cal J}_1,\ldots,{\cal J}_M\subseteq {\cal K}$ denote $M$ disjoint subsets such that $\lambda_k$'s are equal to each other, $k\in{\cal J}_m$, for any $1\leq m\leq M$ and $|{\cal J}_m|\geq 2$; then there generally exist a total number of $\prod_{m=1}^M|{\cal J}_m|!$ optimal decoding orders for problem \eqref{eq.prob_max2}. Here, we can choose any one of them only for the purpose of evaluating the dual function $g(\bm\lambda)$.
\end{remark}

By combining the optimal solutions of $\ell_k^*$'s for the subproblems in \eqref{eq.fk} and $(\bm p^*,\bm r^*)$ for problem \eqref{eq.g0}, the dual function $g(\bm\lambda)$ in \eqref{eq.dual_func} is finally obtained.

\subsection{Finding the Optimal $\bm\lambda^{\rm opt}$ to Maximize $g(\bm \lambda)$ in (D1)}
With $g(\bm \lambda)$ obtained, we solve the dual problem (D1) to obtain the optimal $\bm \lambda^{\rm opt}$ to maximize $g(\bm\lambda)$. Note that $g(\bm \lambda)$ is a convex function but may not be differentiable in general. As a result, problem (D1) can be solved by subgradient based methods such as the ellipsoid method \cite{BoydSlides}, where the subgradient of the objective function $g(\bm \lambda)$ is
\begin{align}\label{eq.subg}
&\left[ r_1^*-{\ell^*_1}/\tilde{T},\ldots,r^*_K-{\ell^*_K}/\tilde{T} \right]^\dagger.
\end{align}

\subsection{Constructing Primal Optimal Solution and Decoding Order for (P1)}\label{sec:cons}

Based on the dual optimal solution $\bm \lambda^{\rm opt}$ to problem (D1), we need to obtain the primal optimal solution $(\mv \ell^{\rm opt}, \mv p^{\rm opt},\mv r^{\rm opt})$ to problem (P1), as well as the optimal decoding order at the BS receiver. Note that under any given $\mv\lambda$, the optimal solution of $\mv \ell^{*}$ and $\mv p^{*}$ to problem \eqref{eq.dual_func} is unique. Therefore, by replacing $\bm \lambda^{\rm opt}$ with $\bm \lambda^{*}$ in Lemma \ref{lem_L} and solving problem \eqref{eq.prob_max2} with $\bm \lambda^{\rm opt}$, one can obtain the primal optimal solution of $\mv \ell^{\rm opt}$ and $\mv p^{\rm opt}$ to problem (P1).

It remains to determine the primal optimal offloading rate $\mv r^{\rm opt}$ and the associated optimal decoding order at the BS for MMSE-SIC. First, consider the case when $\lambda_k^{\rm opt}$'s are different from each other. Let $\mv \pi^{\rm opt}\triangleq [\pi^{\rm opt}(1),\ldots,\pi^{\rm opt}(K)]^\dagger$ denote the permutation such that $\lambda^{\rm opt}_{\pi^{\rm opt}(1)}> \cdots> \lambda^{\rm opt}_{\pi^{\rm opt}(K)} \ge 0$, which then corresponds to the optimal decoding order at the BS receiver for problem (P1). In this case, the primal optimal offloading rate for problem (P1) is obtained as $\bm r^{\rm opt} = [r_{\pi^{\rm opt}(1)}^{\rm opt},\cdots,r_{\pi(K)}^{\rm opt}]^\dagger$ by \eqref{eqn:r:pi:k}.

Next, we consider the case when there exist some $\lambda_k^{\rm opt}$'s that are equal to each other. In this case, it is shown in Remark \ref{remark:1} that the offloading rate $\mv r^*$ and the associated decoding order to problem \eqref{eq.g0} (and thus \eqref{eq.dual_func}) are generally not unique, and thus may not be primal optimal to problem (P1). Therefore, we implement an additional step to construct the primal optimal offloading rate $\mv r^{\rm opt}$ for problem (P1) via proper {\em time-sharing} among different decoding orders as follows. Note that performing time sharing among different corner points at the capacity region of a multiple access channel has been proposed in \cite{Tse98} as a low-complexity scheme to achieve all boundary points of this region. 

In particular, let ${\cal J}_1,\ldots,{\cal J}_M\subseteq {\cal K}$ denote $M$ disjoint subsets such that $\lambda_j^{\rm opt}$'s are identical, $j\in{\cal J}_m$, for any $1\leq m\leq M$ and $|{\cal J}_m|\geq 2$. Define set ${\cal I}\triangleq \{1,\ldots,\prod_{m=1}^M|{\cal J}_m|!\}$. As a result, under the optimal dual solution $\mv \lambda^{\rm opt}$, problem \eqref{eq.g0} (or equivalently \eqref{eq.dual_func}) admits a number of $|\cal I |$ optimal decoding orders, denoted by $\mv \pi^{(1)},\ldots, \mv \pi^{(|\cal I |)}$, and $|\cal I |$ associated optimal offloading rates by \eqref{eqn:r:pi:k}, denoted by $\mv r^{(1)},\ldots,\mv r^{(|\cal I |)}$, where $\mv r^{(i)}=[r_1^{(i)},\ldots,r_K^{(i)}]^\dagger$. Here, the rate tuple $\bm r^{(i)}$ corresponds to one vertex of the polymatroid ${\cal X}(\mv p^{\rm opt})$ for any $i\in\{1,\ldots,|\cal I |\}$; however, the primal optimal rate tuple $\bm r^{\rm opt}$ may lie on a surface of this polymatroid in order to ensure constraint \eqref{eq.rate_cons} in problem (P1). Hence, it is necessary to employ time-sharing among the $|\cal I|$ number of $\bm r^{(i)}$'s. Towards this end, we partition the duration-$T$ block into a total number of $|\cal I |$ time slots. In each slot $i\in \{1,\ldots,|\cal I |\}$ with duration $t^{(i)}$, the $K$ users transmit with rate tuple $\mv r^{(i)}$ and the BS decodes with order $\mv \pi^{(i)}=[\pi^{(i)}(1),\ldots,\pi^{(i)}(K)]^\dagger$. In order to find the optimal time-sharing strategy to solve problem (P1) while satisfying the offloading rate constraints in \eqref{eq.rate_cons}, we obtain the optimal duration $t^{(i)}_{\rm opt}$'s by solving the following feasibility problem:
\begin{subequations}\label{eq.TS_fea}
\begin{align}
{\rm Find} & \quad  t^{(1)},\ldots,t^{(|{\cal I}|)} \\
{\rm s.t. } &\quad
\sum_{i\in{\cal I}} r^{(i)}_k t^{(i)} \geq L_k-\ell_k, \quad \forall k\in{\cal K}\\
&\quad \sum_{i\in{\cal I}} t^{(i)} \leq \tilde{T} \\
& \quad 0\leq t^{(i)}\leq \tilde{T}, \quad \forall i\in{\cal I}.
\end{align}
\end{subequations}
Note that problem \eqref{eq.TS_fea} is a linear program (LP) and can thus be efficiently solved via CVX. Based on the above analysis, we obtain the optimal primal solution of the offloading rate $\mv r^{\rm opt}$ to problem (P1), which is stated in the following proposition.

\begin{proposition}\label{prop2}
When there exist some $\lambda_k^{\rm opt}$'s that are equal to each other, the optimal offloading rate $\mv r^{\rm opt}$ for problem (P1) is obtained by time-sharing among the $|\cal I |$ optimal offloading rates, i.e., $\mv r^{(1)},\ldots,\mv r^{(|\cal I |)}$. In particular, we partition the block into $|\cal I |$ slots, with $t^{(i)}_{\rm opt}$'s denoting the duration of slot $i \in \{1,\ldots,|\cal I |\}$. Then, in each slot $i$, the $K$ users transmit with offloading rate $\mv r^{(i)}$ and the BS receiver decodes with the corresponding order $\mv \pi^{(i)}$.
\end{proposition}

\begin{table}[htp]
\begin{center}
\caption{Algorithm 1 for Solving Problem (P1)}
\hrule \vspace{0.1cm}
\begin{itemize}
\item[a)]
{\bf Initialization:}
Given an ellipsoid ${\cal E}^{(0)}\subseteq{\mathbb R}^{K\times1}$, centered at $\bm\lambda^{(0)}$ and containing the optimal dual solution $\bm\lambda^{\rm opt}$, and set $n=0$.
\item[b)]
{\bf Repeat:}
\begin{itemize}
\item Obtain $\mv \ell^*$ based on \eqref{eq.l_opt}; set the permutation $\bm \pi$ such that $\lambda_{\pi(1)}\geq \cdots\geq \lambda_{\pi(K)}\geq 0$, then obtain $\mv p^*$ by solving problem \eqref{eq.prob_max2}, and get $\mv r^*$ from \eqref{eqn:r:pi:k};
\item Update the ellipsoid ${\cal E}^{(n+1)}$ via the ellipsoid method based on ${\cal E}^{(n)}$ and the subgradient of $g(\bm\lambda)$ given by \eqref{eq.subg}, and set $\bm\lambda^{(n+1)}$ as the center of ellipsoid ${\cal E}^{(n+1)}$;
    \item Set $n \gets n+1$.
 \end{itemize}
\item[c)]
 {\bf Until} the stopping criteria for the ellipsoid method is met.
 \item[d)]
 {\bf Set} $\boldsymbol{\lambda}^{\rm opt} \gets \boldsymbol{\lambda}^{(n)}$.
\item[e)]
{\bf Output}: Obtain $\mv \ell^{\rm{opt}}$ based on \eqref{eq.l_opt} by replacing $\mv\lambda$ with $\mv\lambda^{\rm opt}$, obtain $\mv p^{\rm opt}$ by solving problem \eqref{eq.prob_max2} with $\bm\lambda^{\rm opt}$, and construct the optimal offloading rate $\mv r^{\rm opt}$ and the associated decoding order based on Proposition~\ref{prop2}.
\end{itemize}
\end{center}
\hrule
\end{table}

\begin{figure}
  \centering
  \includegraphics[width = 3.7in]{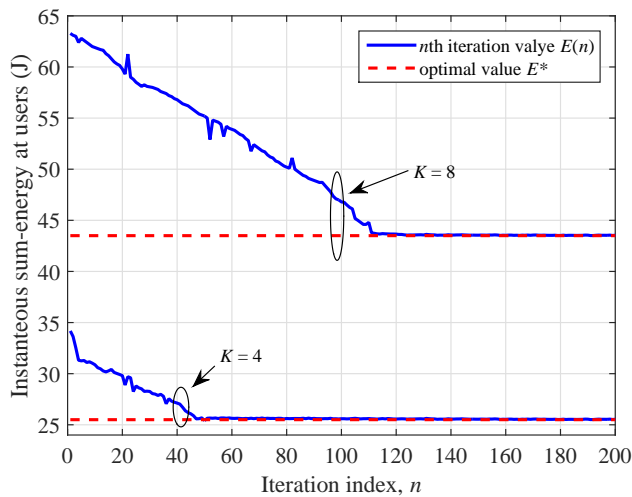}
  \caption{The convergence behavior of Algorithm 1 with the error tolerance $\epsilon=0.01$ and the initial ellipsoid being set as ${\cal E}^{(0)}=\{\bm \lambda\in \mathbb{R}^{K\times1}:\|\bm \lambda\|\leq 20\}$, where $E^*$ and $E(n)$ denote the optimal objective value of problem (P1) and the objective value of problem (P1) after the $n$th iteration in Algorithm 1, respectively.}\label{fig:ConvAlg1}
\end{figure}

In summary, we present Algorithm 1 in Table I to obtain the optimal solution $(\mv \ell^{\rm opt},\mv p^{\rm opt},\mv r^{\rm opt})$ to problem (P1). Note that as problem (P1) is convex, the proposed Algorithm 1 (based on the Lagrange duality method and ellipsoid algorithm\cite{BoydSlides} for solving problem (D1)) is guaranteed to converge to the globally optimal solution. Towards computing the optimal value of problem (P1) with error tolerance $\epsilon$, it takes no more than $2K^2\log(RG/\epsilon)$ iterations for updating the dual variables in Algorithm~1, where $R$ and $G$ denote the radius of the initial ellipsoid ${\cal E}^{(0)}$ and the Lipschitz bound on the objective value of problem (P1), respectively~\cite{BoydSlides}. The fast convergence of Algorithm 1 is corroborated by the numerical results in Fig. \ref{fig:ConvAlg1}, where the block duration is $T=0.3$ sec and the number of task input-bits per user is $L=6\times 10^5$. The remaining parameters are set the same as those in Section V. It is observed that when the number of users is $K=4$, Algorithm~1 yields a solution with desired accuracy within around 50 iterations; while when $K=8$, Algorithm 1 converges within about 120 iterations.

\section{Proposed Solutions to Problem (P2)}\label{sec:binary}
In this section, we first present the BnB based algorithm to find the globally optimal solution to problem (P2.1) (and equivalently (P2)), and then propose two low-complexity algorithms based on the greedy method and convex relaxation, respectively, to find suboptimal solutions with high quality.

\subsection{Optimal Solution Based on BnB algorithm}\label{subsec:BnB}
\begin{figure}
  \centering
  \includegraphics[width = 4.0in]{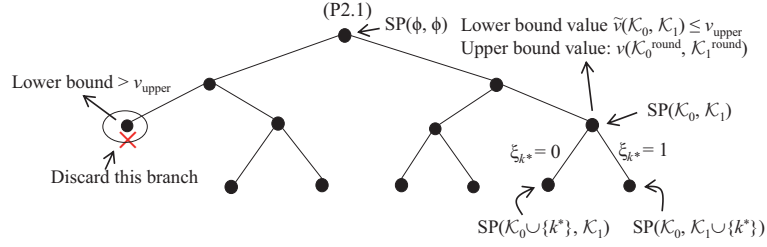}
  \caption{An illustration of the rooted tree for the BnB algorithm to solve the mixed Boolean convex problem (P2.1).} \label{fig.BnB_tree}
\end{figure}

In this subsection, we propose a BnB based algorithm to obtain the globally optimal solution to problem (P2.1). BnB is a widely adopted approach to optimally solve discrete and combinational optimization problems via the so-called state space search \cite{Jens99}. In the BnB algorithm, the set of candidate solutions of the integer variables (e.g., $\xi_k$'s in problem (P2.1) of our interests) is formed as a rooted tree, in which the root corresponds to the full set, and each branch represents a subset of the candidate solutions. Next, the BnB algorithm successively checks branches of this tree. For each branch, a corresponding problem is constructed with certain given variables, based on which we can accordingly obtain the upper and lower bounds on the optimal value to the original problem. If the lower bound of a branch (for a minimization problem) cannot lead to a better solution than the best solution found so far, the corresponding branch will be discarded. After checking all the branches, the BnB algorithm is able to find the globally optimal solution to the discrete or combinational optimization problem. Please refer to \cite{Jens99,BoydSlides_BnB} and references therein for more detailed description about the BnB algorithm.

To use the BnB algorithm for finding the globally optimal solution to problem (P2.1), we first define the following weighted sum-energy minimization problem under given offloading/local computing decisions at certain users:
\begin{subequations}\label{prob.sq}
\begin{align}
{\rm SP}({\cal K}_0,{\cal K}_1):~ \min_{\mv \xi, \mv p, \mv r} &~~  \sum_{k=1}^K \alpha_k\left(\frac{\zeta_kC_k^3L_k^3(1-\xi_k)^3}{T^2}+p_k \tilde{T} \right)\\
  {\rm s.t.}~~& \mv r \in {\cal X}(\mv p) \\
 &  r_k\geq \xi_k L_k/\tilde{T},~~ \forall k\in{\cal K}\\
 & p_k \geq 0, ~~ \forall k\in{\cal K}\\
&  \xi_k =  0,~~ \forall k\in{\cal K}_0\\
&  \xi_k =  1,~~ \forall k\in{\cal K}_1\\
&  \xi_k \in\{0,1\},~~ \forall k\in{\cal K}\setminus({\cal K}_0\cup{\cal K}_1),
\end{align}
\end{subequations}
where ${\cal K}_0\subseteq {\cal K}$ and ${\cal K}_1\subseteq{\cal K}$ denote two disjoint sets of users with ${\cal K}_0\cap {\cal K}_1=\emptyset$, such that each user in ${\cal K}_0$ chooses to locally compute its tasks and each user in ${\cal K}_1$ chooses to offload the tasks to the BS, as defined in (\ref{prob.sq}e) and (\ref{prob.sq}f), respectively. Notice that in the case with ${\cal K}_0\cup {\cal K}_1\neq {\cal K}$, problem ${\rm SP}({\cal K}_0,{\cal K}_1)$ is still a mixed Boolean convex problem. However, when ${\cal K}_0\cup {\cal K}_1={\cal K}$, ${\rm SP}({\cal K}_0,{\cal K}_1)$ becomes a convex optimization problem that can be efficiently solved, similarly by the proposed Algorithm 1 in Table I.

With problem ${\rm SP}({\cal K}_0,{\cal K}_1)$ at hand, the BnB algorithm for solving problem (P2.1) is implemented by iteratively constructing a rooted tree with each vertex corresponding to a problem ${\rm SP}({\cal K}_0,{\cal K}_1)$ with given ${\cal K}_0$ and ${\cal K}_1$, as depicted by Fig.~\ref{fig.BnB_tree}. Initially, we have the root of this tree as problem ${\rm SP}(\emptyset,\emptyset)$ (or equivalently problem (P2.1)). At each iteration, we check the estimated upper and lower bound values of problem ${\rm SP}({\cal K}_0,{\cal K}_1)$'s at all the outer vertices of this tree, where the best upper and lower bound values are denoted as $v_{\rm upper}$ and $v_{\rm lower}$, respectively. For the vertices whose lower bound values are larger than the current best upper bound value $v_{\rm upper}$, we discard these vertices, since all feasible solutions corresponding to these vertices are worse than the current best solution. As for the remaining outer vertices, each one is further branched into two branches (or two new outer vertices), which correspond to the two problems ${\rm SP}({\cal K}_0\cup\{k^*\},{\cal K}_1)$ and ${\rm SP}({\cal K}_0,{\cal K}_1\cup\{k^*\})$, respectively, where user $k^*$ is properly selected from set ${\cal K}\setminus({\cal K}_0\cup{\cal K}_1)$. Then, we estimate the upper and lower bound values of these new branches, and update the best bound values $v_{\rm upper}$ and $v_{\rm lower}$. The iteration terminates until the difference between the best upper and lower bound values (i.e., $v_{\rm upper}-v_{\rm lower}$) is smaller than a given error tolerance. In the following, we present the two main steps of the BnB algorithm, i.e., the bounding procedure that estimates the upper and lower bound values for problem ${\rm SP}({\cal K}_0,{\cal K}_1)$ at each vertex (or branch) of this tree, and the branching procedure that chooses the outer vertices to be branched at each iteration.

\subsubsection{Bounding}
Consider one particular vertex corresponding to problem ${\rm SP}({\cal K}_0,{\cal K}_1)$. We describe how to estimate upper and lower bound values for problem ${\rm SP}({\cal K}_0,{\cal K}_1)$. To this end, we denote the convex relaxation problem of ${\rm SP}({\cal K}_0,{\cal K}_1)$ as $\widetilde{\rm SP}({\cal K}_0,{\cal K}_1)$, which is obtained based on ${\rm SP}({\cal K}_0,{\cal K}_1)$ by substituting the constraints (\ref{prob.sq}g) with $0\leq \xi_k \leq 1$  for all $k\in{\cal K}\setminus({\cal K}_0\cup{\cal K}_1)$. Since that $\widetilde{\rm SP}({\cal K}_0,{\cal K}_1)$ is a convex optimization problem, it can be similarly solved by Algorithm~1 in Table I. Let $\tilde{v}({\cal K}_0,{\cal K}_1)$ and $(\tilde{\bm \xi},\tilde{\bm p},\tilde{\bm r})$ denote the optimal value and the corresponding optimal solution of problem $\widetilde{\rm SP}({\cal K}_0,{\cal K}_1)$, respectively. Since that the feasible region of problem ${\rm SP}({\cal K}_0,{\cal K}_1)$ is contained inside that of $\widetilde{\rm SP}({\cal K}_0,{\cal K}_1)$, $\tilde{v}({\cal K}_0,{\cal K}_1)$ is a lower bound value for the optimal value of ${\rm SP}({\cal K}_0,{\cal K}_1)$. As for an upper bound for the optimal value of problem ${\rm SP}({\cal K}_0,{\cal K}_1)$, we implement a rounding procedure based on the optimal solution of $\tilde{\bm \xi}$ to problem $\widetilde{\rm SP}({\cal K}_0,{\cal K}_1)$, such that the user sets ${\cal K}^{\rm round}_0$ and ${\cal K}^{\rm round}_1$ are obtained as
\begin{align}\label{eq.rounding}
\begin{cases}
{\cal K}^{\rm round}_0 \triangleq \{k:\;0\leq \tilde{\xi}_k<0.5,\; k\in{\cal K}\}\\
{\cal K}^{\rm round}_1\triangleq \{k:\;0.5\leq \tilde{\xi}_k\leq 1,\; k\in{\cal K}\}.
\end{cases}
\end{align}
Note that ${\cal K}^{\rm round}_0\cup{\cal K}^{\rm round}_1={\cal K}$. As a result, for problem ${\rm SP}({\cal K}^{\rm round}_0,{\cal K}^{\rm round}_1)$, the $K$ users' offloading/local computing decisions are fixed to $\xi_k=0$ for all $k\in{\cal K}_0^{\rm round}$ and $\xi_k=1$ for all $k\in{\cal K}_1^{\rm round}$, respectively. The optimal value to problem ${\rm SP}({\cal K}^{\rm round}_0,{\cal K}^{\rm round}_1)$, denoted by $v({\cal K}^{\rm round}_0,{\cal K}^{\rm round}_1)$, can be efficiently obtained by Algorithm 1 in Table I again. It is evident that $v({\cal K}^{\rm round}_0,{\cal K}^{\rm round}_1)$ serves as an upper bound value for the optimal value of problem ${\rm SP}({\cal K}_0,{\cal K}_1)$.

\subsubsection{Branching}
At each iteration, we select the remaining outer vertices each corresponding to problem ${\rm SP}({\cal K}_0,{\cal K}_1)$ with $\tilde{v}({\cal K}_0,{\cal K}_1)\leq v_{\rm upper}$. Based on the optimal solution of $\tilde{\xi}_i$'s for problem $\widetilde{\rm SP}({\cal K}_0,{\cal K}_1)$, we select user $k^*$ for branching as the most ``undecided'' (between local computing and offloading) user from set ${\cal K}\setminus({\cal K}_0\cup{\cal K}_1)$, i.e.,
\begin{align}\label{eq.branch}
k^* \triangleq \argmin_{k\in{\cal K}\setminus({\cal K}_0\cup{\cal K}_1)}~|\tilde{\xi}_{k}-1/2|.
\end{align}
Then problem ${\rm SP}({\cal K}_0,{\cal K}_1)$ is branched into the two problems ${\rm SP}({\cal K}_0\cup\{k^*\},{\cal K}_1)$ and ${\rm SP}({\cal K}_0,{\cal K}_1\cup\{k^*\})$.

With the above bounding and branching procedures, the BnB algorithm can be efficiently used for solving problem (P2.1). It is established from \cite{BoydSlides_BnB} that the proposed BnB algorithm is guaranteed to find the globally optimal solution to problem (P2.1) (and equivalently (P2)). Nevertheless, such an algorithm requires checking the upper and lower bounds for $2^K$ vertices by solving a total of $2^{K+1}$ problems (i.e., $2^K$ problems of $\widetilde{\rm SP}({\cal K}_0,{\cal K}_1)$ and $2^K$ problems of ${\rm SP}({\cal K}^{\rm round}_0,{\cal K}^{\rm round}_1)$) at the worst case. Therefore, this BnB algorithm generally has very high implementation complexity, especially when the number $K$ of users is sufficiently large. For facilitating practical implementation, it calls for low-complexity algorithm designs to efficiently solve problem (P2.1), as will be developed in the following two subsections.

\subsection{Suboptimal Solution Based on Greedy Strategy}\label{subsec:greedy}
In this subsection, we present a low-complexity solution to problem (P2.1) based on the greedy strategy. In this design, similarly as in Section~\ref{subsec:BnB}, we denote the two disjoint user sets as ${\cal K}^{\rm gre}_0$ and ${\cal K}^{\rm gre}_1$ with ${\cal K}_0^{\rm gre}\cap{\cal K}_1^{\rm gre}=\emptyset$, respectively, such that each user $ k\in {\cal K}^{\rm gre}_0$ performs local computing and each user $k\in {\cal K}^{\rm gre}_1$ performs offloading, respectively. Differently, here we restrain the user sets such that ${\cal K}^{\rm gre}_0\cup {\cal K}^{\rm gre}_1={\cal K}$.

The proposed greedy algorithm is implemented in an iterative manner. Initially, we set ${\cal K}^{\rm gre}_0={\cal K}$ and ${\cal K}^{\rm gre}_1=\emptyset$, and the value $v^{(0)*}$ is set as the optimal value to problem ${\rm SP}({\cal K},\emptyset)$. Then, in each iteration $n\geq 1$, we allow each user $k\in{\cal K}_0^{\rm gre}$ to perform computation offloading, and accordingly compute the weighted sum-energy consumption, by solving the convex optimization problem ${\rm SP}({\cal K}^{\rm gre}_0\setminus \{k\},{\cal K}^{\rm gre}_1\cup \{k\})$, for which the optimal value is denoted as $v^{(n)}_k$. Then the user achieving the minimum weighted sum-energy is denoted as $k^{\rm opt} \triangleq \argmin_{k\in{\cal K}^{\rm gre}_0} v^{(n)}_k$. Accordingly, we have $v^{(n)*}=v^{(n)}_{k^{\rm opt}}$. If the value $v^{(n)*}$ is smaller than $v^{(n-1)*}$ in the previous iteration $(n-1)$, then we update ${\cal K}^{\rm gre}_0\gets{\cal K}^{\rm gre}_{0}\setminus \{k^{\rm opt}\}$ and ${\cal K}^{\rm gre}_{1}\gets{\cal K}^{\rm gre}_{1}\cup \{ k^{\rm opt}\}$. If $v^{(n)*}\geq v^{(n-1)*}$ or $n=K$, then this algorithm terminates.

Note that for the greedy based solution, a total of $(K-n+1)$ convex optimization problems ${\rm SP}({\cal K}^{\rm gre}_0,{\cal K}^{\rm gre}_1)$, are required to be solved at each iteration $n\in\{1,\ldots,K\}$, and it requires $K$ iterations at the worst case. Therefore, at the worst case, a total of $K(K+1)/2$ convex problems ${\rm SP}({\cal K}^{\rm gre}_0,{\cal K}^{\rm gre}_1)$ are solved in the proposed greedy algorithm. As compared to the BnB algorithm in Section~\ref{subsec:BnB}, the greedy based algorithm has a significantly reduced complexity, but achieves a close-to-optimal performance, as will be shown in numerical results later.

\subsection{Suboptimal Solution Base on Convex Relaxation}
In this subsection, we proposed another suboptimal solution to problem (P2.1) based on convex relaxation \cite{BoydSlides_BnB}.

First, the Boolean constraint $\xi_k\in\{0,1\}$ in problem (P2.1) is relaxed into the continuous constraint $0\leq \xi_k\leq 1$ for any $k\in{\cal K}$. Accordingly, the mixed Boolean convex problem (P2.1) is relaxed as a convex optimization problem $\widetilde{\rm SP}({\cal K}_0, {\cal K}_1)$ in \eqref{prob.sq} with ${\cal K}_0={\cal K}_1=\emptyset$ (or (P1) equivalently). Next, the rounding operation in \eqref{eq.rounding} is employed to get the corresponding user sets ${\cal K}^{\rm rel}_0$ and ${\cal K}^{\rm rel}_1$. Finally, the suboptimal solution is obtained by solving the convex problem ${\rm SP}({\cal K}^{\rm rel}_0,{\cal K}^{\rm rel}_1)$.

Note that, in order to obtain the suboptimal solution to problem (P2.1) based on the convex relaxation, we only need to solve two convex problems (i.e., $\widetilde{\rm SP}(\emptyset,\emptyset)$ and ${\rm SP}({\cal K}^{\rm rel}_0,{\cal K}^{\rm rel}_1)$) by Algorithm~1. As compared to the greedy algorithm in Section~\ref{subsec:greedy}, the convex relaxation based solution has a significantly reduced computational complexity.

\section{Numerical Results}\label{sec:numerical}
In this section, we provide numerical results to validate the performance of the proposed NOMA-based offloading designs in the multiuser MEC system, as compared with other benchmark schemes in the following.

\begin{enumerate}
\item{\em Local computing only:} All the $K$ users execute their tasks via local computing only. This scheme corresponds to solving problem (P1) or problem (P2) by setting $\ell_k=L_k$, $k\in{\cal K}$. The resultant weighted sum-energy consumption at the users can be expressed in closed-form as $\sum_{k=1}^K \alpha_k\zeta_kC_k^3L_k^3/T^2$. This scheme applies for both partial and binary offloading cases.

\item{\em Full offloading only:} All the $K$ users execute their computation tasks by offloading all the task input-bits to the BS simultaneously. This scheme corresponds to solving problem (P1) by setting $\ell_k=0$, $k\in{\cal K}$, via Algorithm 1 in Table I. This scheme applies for both partial and binary offloading cases.

\item{\em OMA-based offloading \cite{You17}:} The $K$ users adopt a TDMA protocol for computation offloading, where the duration-$T$ block is partitioned into $K$ slots. For the case with partial offloading, user $k$ offloads part of its computation input-bits to the BS in the $k$th time slot for any $k\in{\cal K}$. We jointly optimize the time slot allocation among the users, the transmit power and the number of task input-bits for each user's offloading, and the local CPU frequency per user, in order to minimize the weighted sum-energy consumption at the $K$ users. For the case with binary offloading, targeting on the same objective, we jointly optimize the users' offloading decisions, the local CPU frequencies for the users that perform local computing, and the time allocation among the other users that perform offloading, as well as the transmit powers and rates for the offloading users. The convex optimization technique is employed to solve the weighted sum-energy minimization problem for the case with partial offloading \cite{You17}, and the BnB algorithm is used for solving the problem for the case with binary offloading.
\end{enumerate}

In the simulations, we set the system bandwidth for computation offloading as $B=2$ MHz and the noise power spectrum density (PSD) at the BS receiver as $N_0=-174$ dBm/Hz. We consider the Rayleigh fading channel model, where the channel vector between each user $k\in{\cal K}$ and the BS is established as $\bm h_k = \sqrt{ G_0\Big(\frac{d_k}{d_0}\Big)^{-\theta}} \boldsymbol{\bar{h}}_k$, where $G_0=-40$ dB corresponds to the path loss at a reference distance of $d_0=1$ meter, $d_k$ denotes the distance between user $k$ and the BS, $\theta=3.5$ denotes the path loss exponent, and $\bm{\bar h}_k$ is a CSCG random vector with $\bm{\bar h}_k\sim {\cal CN}(0,\bm I)$. For the local computing at each user $k\in\mathcal K$, we set $C_k=4\times10^3$ cycles/bit and $\zeta_k=10^{-28}$\cite{You17}. The numbers of task input-bits at different users are set to be identical, i.e., $L=L_k$, $k\in{\cal K}$. In addition, the transmission time per block is set to be identical to the block length, i.e., $\tilde{T}=0.9T$.

\begin{figure}
  \centering
  \includegraphics[width = 3.7in]{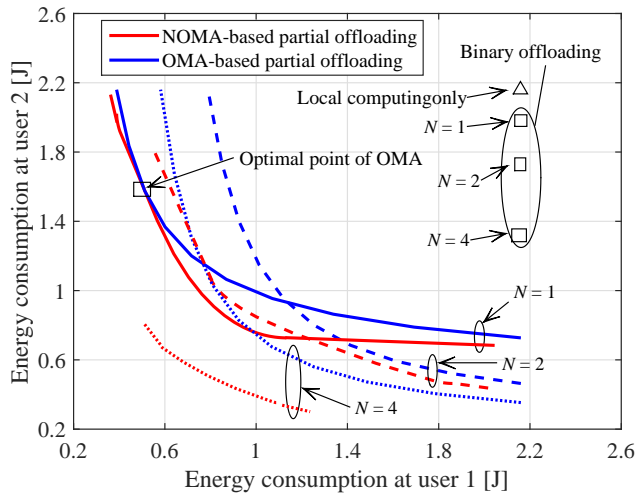}
  \caption{Energy consumption tradeoff between two users.} \label{fig.energy_region}
\end{figure}

First, we consider the basic scenario with a BS equipped with different antennas serving $K = 2$ users. The number of task input-bits at each user is set as $L=5\times 10^5$ bits, the block duration is $T=0.1$ sec. The distances between the BS and the two users are set as $d_1=400$ and $d_2=300$ meters. Under different values of the antenna number $N$ at the BS and by considering one channel realization for each $N$, Fig.~\ref{fig.energy_region} shows the energy consumption tradeoff between the two users, which is obtained by exhausting all the nonnegative user weights $\alpha_1$ and $\alpha_2$ with $\alpha_1+\alpha_2=1$. Note that the two users' energy consumptions by the full-offloading-only scheme are much larger than those by the local-computing-only scheme, and thus are not shown in this figure. It is observed that the NOMA-based partial offloading generally achieves lower energy consumption at both users than the other two benchmark schemes, and the performance gain becomes more significant as $N$ grows. This validates the effectiveness of our proposed NOMA-based partial offloading schemes in minimizing the users' energy consumption while satisfying the computation latency constraints. Note that in the single-BS-antenna case ($N=1$) with $\alpha_1 = 0.35$ under our setup, the OMA-based partial offloading scheme is observed to achieve the same energy-saving performance as the NOMA-based counterpart. This is consistent with the fact that in the single-antenna scenario, the OMA transmission is able to achieve one point at the Pareto boundary of the capacity region for the multiple access channel (that is achievable by NOMA) in a time-sharing manner\cite{VTC17}. Under our setup, this point is achieved by the two users time-sharing the offloading wireless channel with portions $\alpha_1=0.35$ and $\alpha_2=1-\alpha_1=0.65$, respectively. For binary offloading, it is observed that energy consumption of user 2 is reduced as compared to that in the local computing only scheme. This is because user 2 chooses offloading its task in this case.

Next, we consider a more general scenario where the BS is equipped with $N=4$ antennas to serve a total of $K$ users. Assume that the distance between each user $k\in{\cal K}$ and the BS follows a uniform distribution within $[100,400]$ meters. The energy weights for different users are set to be identical as $\alpha_k=1$, $k\in{\cal K}$. Then our objective corresponds to minimizing the sum energy consumption at all the $K$ users. The results are obtained by averaging over 500 randomized channel realizations.

\subsection{The Case with Partial Offloading}

\begin{figure}
  \centering
  \includegraphics[width = 3.7in]{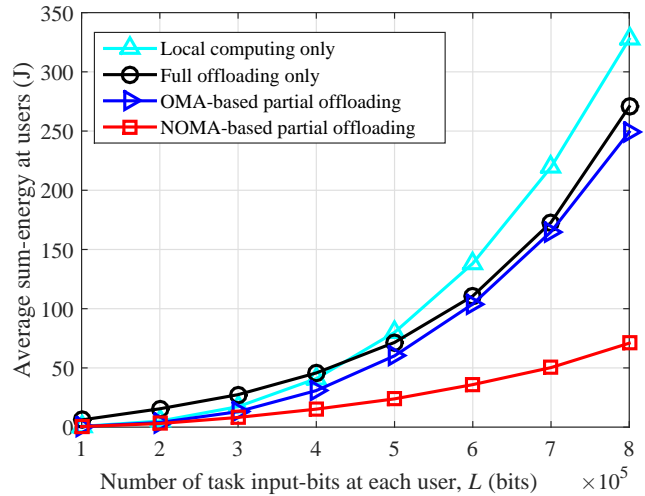}
  \caption{The average sum-energy consumption at the users versus the number of task input-bits $L$ at each user.} \label{fig.vs_size}
\end{figure}


Fig.~\ref{fig.vs_size} shows the average sum-energy consumption at these users versus the number $L$ of task input-bits for each user, where $T=0.2$ sec and $K=4$. It is observed that the averaged sum-energy consumption by all the schemes increases as $L$ increases. The proposed NOMA-based partial offloading scheme is observed to achieve the smallest energy performance among all the schemes. Compared with the OMA-based partial offloading scheme, substantial gains are observed by the proposed NOMA-based one, especially when $L$ becomes large. It is also observed that the OMA-based scheme outperforms both the local-computing-only and full-offloading-only schemes. The benchmark local-computing-only scheme outperforms the full-offloading-only scheme, but performs inferior when $L$ becomes large, e.g., $L\geq 4\times 10^5$ bits.

\begin{figure}
  \centering
  \includegraphics[width = 3.7in]{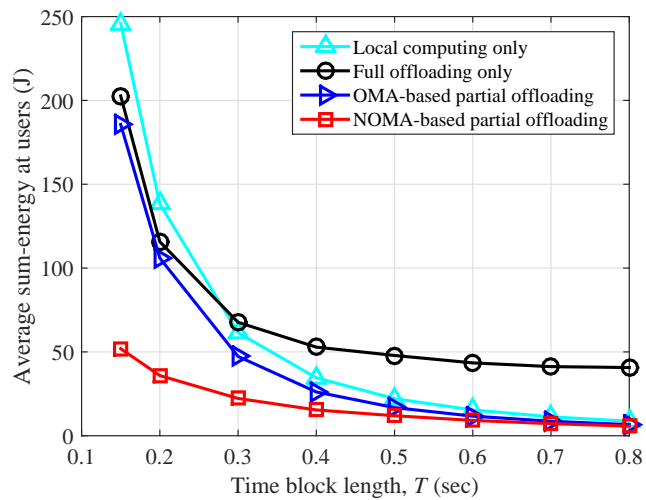}
  \caption{The average sum-energy consumption at the users versus the duration $T$ of the time block.} \label{fig.vs_TimeBlock}
\end{figure}

Fig. \ref{fig.vs_TimeBlock} shows the average sum-energy consumption at these users under different time block duration $T$, where $L=6\times 10^5$ bits and $K=4$. In general, we have similar observations as in Fig.~\ref{fig.vs_size}. Particularly, the NOMA-based partial offloading scheme is observed to achieve a much better performance than the OMA one when $T$ becomes small.

\begin{figure}
  \centering
  \includegraphics[width = 3.7in]{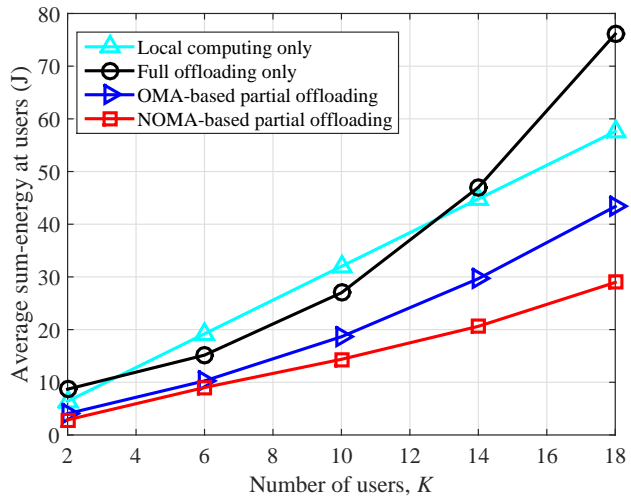}
  \caption{The average sum-energy consumption at the users versus the user number $K$.} \label{fig.vs_K}
\end{figure}

Fig. \ref{fig.vs_K} shows the average sum-energy consumption at these users under different numbers of users, where $T=0.5$~sec and $L=5\times 10^5$ bits. It is observed that the proposed NOMA-based partial offloading scheme achieves substantial gains over the benchmark schemes, especially when the number $K$ of users becomes large. By combining this with Figs.~\ref{fig.vs_size} and~\ref{fig.vs_TimeBlock}, it suggests that the NOMA-based partial offloading scheme becomes increasingly essential in energy saving when the offloading resources become stringent among multiple users.

\begin{figure}
 \centering
  \includegraphics[width = 3.7in]{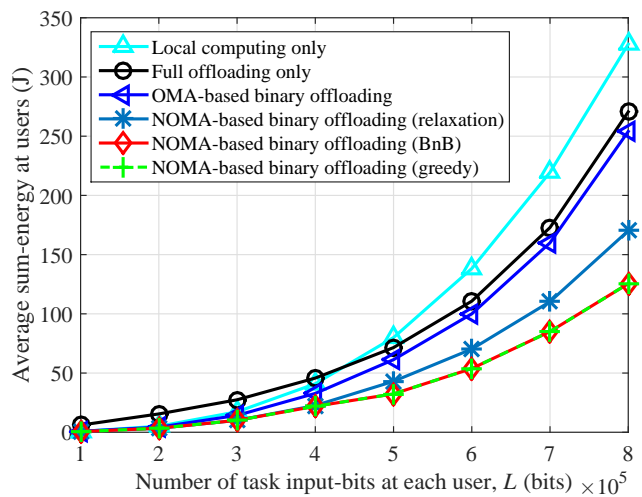}
  \caption{The average sum-energy consumption at the users versus the number of task input-bits $L$ at each user.} \label{fig.vs_size_Binary}
\end{figure}

\subsection{The Case With Binary Offloading}

Fig.~\ref{fig.vs_size_Binary} shows the average sum-energy consumption at these users versus the number of task input-bits $L$ for each user, where $T=0.2$ sec and $K=4$. It is observed that the propose NOMA-based binary offloading schemes (under the three approaches of BnB, greedy, and convex relaxation) all outperform the other three benchmark schemes, and the performance gain increases as $L$ grows. It is also observed that the proposed greedy algorithm achieves a close-to-optimal performance as the BnB algorithm. As the greedy algorithm has a much lower complexity than the BnB algorithm, this indicates that the greedy algorithm is very efficient in practical implementation. Furthermore, the OMA-based binary offloading scheme is observed to achieve a better performance than the local-computing-only and the full-offloading-only schemes.

\begin{figure}
  \centering
  \includegraphics[width = 3.7in]{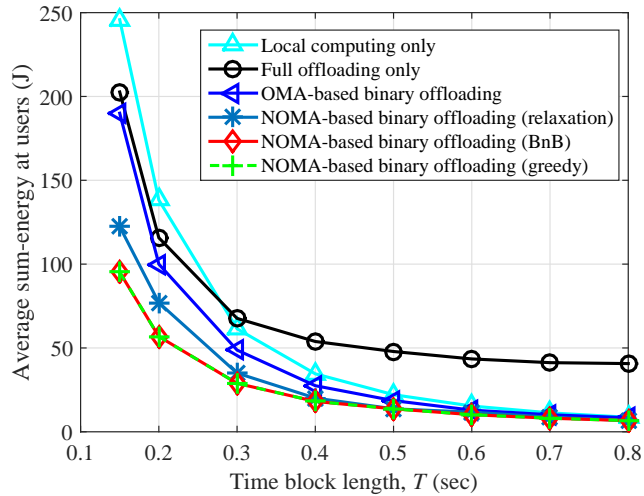}
  \caption{The average sum-energy consumption at the users versus the duration $T$ of the time block.} \label{fig.vs_TimeBlock_Binary}
\end{figure}

Fig. \ref{fig.vs_TimeBlock_Binary} shows the average sum-energy consumption at these users under different time block duration $T$ for the binary offloading case, where $L=6\times 10^5$ bits and $K=4$. In general, we have similar observations as in Fig.~\ref{fig.vs_size_Binary}. At small $T$ values, the proposed NOMA-based binary offloading schemes achieve substantial gains over the OMA-based binary offloading scheme. 

\begin{figure}
  \centering
  \includegraphics[width = 3.7in]{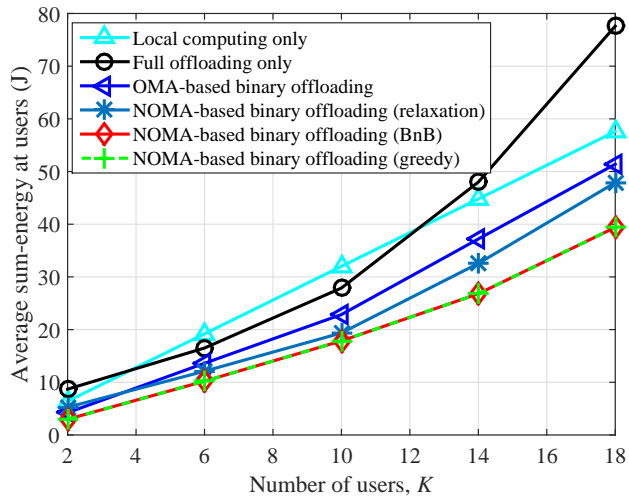}
  \caption{The average sum-energy consumption at the users versus the user number $K$.} \label{fig.vs_K_Binary}
\end{figure}

Fig.~\ref{fig.vs_K_Binary} shows the average sum-energy consumption at users for different numbers of users, where the parameters are set the same as those in Fig.~\ref{fig.vs_K}. The proposed greedy algorithm is observed to achieve a close-to-optimal performance as the BnB algorithm. As the number $K$ of users increases, the NOMA based scheme outperforms the benchmark OMA-based scheme and the performance gain becomes significant when $K$ becomes sufficiently large. Together with Figs.~\ref{fig.vs_size_Binary} and \ref{fig.vs_TimeBlock_Binary}, this indicates the benefits of NOMA-based multiuser offloading in energy saving when the communication resources (e.g., offloading time) for each user become scarce.


\begin{figure}
  \centering
  \includegraphics[width = 3.7in]{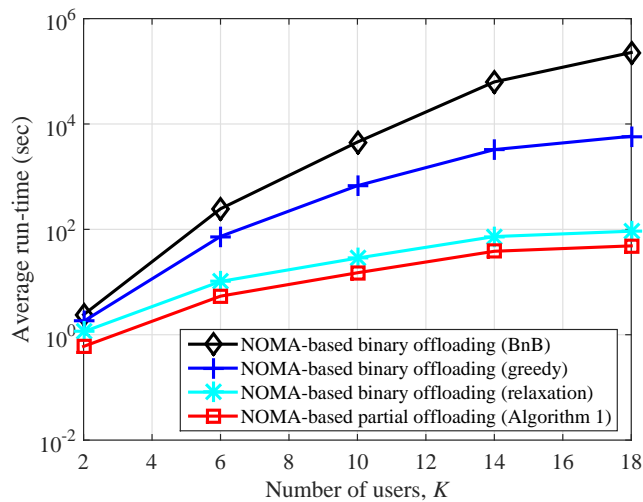}
\caption{The average run-time comparison of the proposed algorithms under different numbers of users, where the error tolerance is $\epsilon=0.01$ and the initial ellipsoid is ${\cal E}^{(0)}=\{\bm \lambda\in \mathbb{R}^{K\times1}:\|\bm \lambda\|\leq 20\}$.}\label{fig:run-time}
\end{figure}

Next, we compare the complexity of different algorithms proposed in this paper, including Algorithm 1 for solving problem (P1), as well as the BnB, greedy strategy, and convex relation based algorithms for solving problem (P2.1). Fig.~\ref{fig:run-time} shows the average run-time of these algorithms under different numbers of users, where $T=0.3$ sec and $L=6\times 10^5$. The results are obtained by using MATLAB R2015a with a Dell Precision 3620 Desktop (Windows 7 Professional service pack 1, Intel(R) Xeon(R) CPU E3-1220 v5 @3.00GHz, RAM 16.0GB). It is observed that the average run-time for all the four algorithms increases with $K$, which is due to the fact that a larger $K$ value leads to a large number of decision variables in both problems (P1) and (P2). In particular, Algorithm 1 is observed to achieve the smallest run-time value among the four algorithms. This is because the other three algorithms involve Algorithm 1 as their sub-procedure, as shown in Section IV. For the NOMA-based binary offloading scheme, the BnB based algorithm is observed to consume the longest run time to find the optimal solution to problem (P2) among the three schemes, and the greedy method is observed to consume longer run time than the convex relaxation based one. More specifically, in the case with $K=18$, the complexity of the greedy method is only around $1/80$ of that of the BnB algorithm, while that of the convex relaxation based algorithm is about $1/90$ of that of the greedy method. Such results are consistent with the computational complexity analysis in Section IV, which show that the convex relaxation, greedy strategy, and BnB based algorithms involve a total number of 2, $K(K+1)/2$, and $2^{K+1}$ of convex problems (i.e., ${\rm SP}({\cal K}_0,{\cal K}_1)$ and/or $\widetilde{\rm SP}({\cal K}_0,{\cal K}_1)$) to be solved, respectively.

\section{Conclusion}\label{sec:conclusion}
This paper proposes to explore the full potential of computation offloading via multi-antenna NOMA for improving the performance of multiuser MEC systems. By considering both cases with partial and binary offloading, we minimize the weighted sum-energy by jointly optimizing the users' CPU frequencies for local computing, their transmit powers and rates for computation offloading, as well as the BS's decoding order for MMSE-SIC. We propose efficient algorithms to solve the weighted sum-energy minimization problems under both cases. Numerical results show that the proposed NOMA-based offloading schemes enjoy substantial weighted sum-energy consumption performance gains (or weighted sum-energy consumption reduction), as compared to the benchmark schemes such as the conventional OMA-based offloading designs.

There remains a lot of interesting problems, which are unaddressed in this paper but worthy of investigation in future work. For example, it is interesting to consider a mixture scenario where the users with partial offloading and those with binary offloading may coexist within the same MEC network. How to extend our proposed designs (e.g., the BnB, greedy, and convex relaxation based algorithms) to this scenario is worth studying. Next, computation users may also coexist with communication users with different quality of service (QoS) requirements. How to design an energy-efficient network by considering these constraints as well as their fairness issues is an important but challenging problem. Furthermore, solely using the MEC serves at the BSs may not be sufficient to ensure the QoS requirements of all users, especially when the number of users is large and/or the computation loads are heavy. How to integrate the distributed edge computing with the centralized cloud computing to deal with such challenges is another interesting direction to pursue.

\section*{Appendices}

\subsection{Proof of Proposition \ref{prop.NP-hard}}
Consider one special case of problem (P2.1), in which the wireless channels of the $K$ users are orthogonal to each other, with $K\leq N$. In this case, at the optimality of problem (P2.1), it must hold that $r_k=\xi_kL_k/\tilde{T}$ and $p_k=\xi_k\frac{2^{L_k/(\tilde{T}B)}-1}{\|\bm h_k\|^2}$, $\forall k\in{\cal K}$. By substituting $r_k$ and $p_k$, and noticing that $(1-\xi_k)^3=1-\xi_k$, $\forall \xi_k\in\{0,1\}, k\in{\cal K}$, it follows that problem (P2.1) can be re-expressed as the following integer program:
\begin{subequations}\label{prob.IP}
\begin{align}
\min_{\bm \xi}~& c_0+\bm m^H\bm \xi\\
{\rm s.t. }~&~\xi_k\in\{0,1\},~~\forall k\in{\cal K}.
\end{align}
\end{subequations}
where $c_0\triangleq \sum_{k=1}^K\frac{\alpha_k\zeta_kC_k^3}{T^2}$ and $\bm m \triangleq \left[\frac{(2^{L_1/(\tilde{T}B)}-1)\alpha_1\tilde{T}}{\|\bm h_1\|^2}-\frac{\alpha_1\zeta_1C_1^3}{T^2},\ldots,\frac{(2^{L_K/(\tilde{T}B)}-1)\alpha_K\tilde{T}}{\|\bm h_K\|^2}-\frac{\alpha_K\zeta_KC_K^3}{T^2}\right]^\dagger$. Note that the integer program \eqref{prob.IP} corresponds to the following well-established NP-complete decision \cite[Section 2.10]{Wig06}: Given a polytope in ${\mathbb R}^K$ (by its bounding hyperplanes), does it contain an integer point?

It thus follows that problem \eqref{prob.IP} is an NP-hard problem. Notice that in general cases when the wireless channels of the $K$ users are not orthogonal, problem (P2.1) is actually more difficult than problem \eqref{prob.IP}. Therefore, problem (P2.1) (or its equivalent problem (P2)) is an NP-hard problem. As a result, Proposition \ref{prop.NP-hard} is proved.

\subsection{Proof of Lemma \ref{lem_L}}
Given ${\bm \lambda}$, we solve problem \eqref{eq.fk} for each user $k\in{\cal K}$. The objective function of problem \eqref{eq.fk} is convex with respect to $\ell_k$ and the constraint $0\leq \ell_k\leq L_k$ is linear. Therefore, \eqref{eq.fk} is a convex optimization problem and satisfies the Slater's condition. The Lagrangian of the $k$th subproblem in \eqref{eq.fk} is given by
\begin{align}\label{eq.La_Lk}
 {\cal L}_k(\ell_k,\bar{\beta}_k,\underline{\beta}_k) =\frac{\alpha_k\zeta_kC_k^3}{T^2}(L_k-\ell_k)^3+\frac{\lambda_k}{B\tilde{T}}\ell_k - \bar{\beta}_k \ell_k - \underline{\beta}_k(L_k-\ell_k),
\end{align}
where $\bar{\beta}_k\geq 0$ and $\underline{\beta}_k\geq 0$ are the Lagrange multipliers associated with $\ell_k\geq 0$ and $\ell_k\leq L_k$, respectively. Let $(\bar{\beta}_k^{*},\underline{\beta}_k^{*})$ denote the optimal dual solution, and $\ell_k^*$ denote the optimal primal solution. Based on the KKT conditions\cite{BoydBook}, it follows that
\begin{subequations} \label{eq.Lk_kkt}
\begin{align}
& L_k\geq \ell_k^{*},~~\ell_k^{*}\geq 0,~~\bar{\beta}_k^{*}\geq 0,~~\underline{\beta}_k^{*}\geq 0 \\
& \bar{\beta}_k^{*}\ell_k^{*}=0,~~\underline{\beta}^{*}_k(L_k-\ell_k^{*})=0 \\
&  -\frac{3\alpha_k\zeta_kC_k^3}{T^2}(L_k-\ell^{*}_k)^2+\frac{\lambda_k}{B\tilde{T}}-\bar{\beta}^*_k+\underline{\beta}^*_k=0,
\end{align}
\end{subequations}
where (\ref{eq.Lk_kkt}a) denotes the primal and dual feasibility conditions, (\ref{eq.Lk_kkt}b) corresponds to the complementary slackness conditions, and the LHS of (\ref{eq.Lk_kkt}c) is the first-order derivative of ${\cal L}_k(\ell_k,\bar{\beta}_k,\underline{\beta}_k)$ with respect to $\ell_k$. Based on (\ref{eq.Lk_kkt}b), it follows that $\bar{\beta}^*_k=\underline{\beta}^*_k=0$ for $0< \ell^*_k < L_k$. After some simple manipulations, we have
$\ell_k^* =
\left(L_k-\sqrt{\frac{T\lambda_k}{3B\alpha_k\zeta_kC_k^3}}\right)^+$, $\forall k\in {\cal K}$.


 \ifCLASSOPTIONcaptionsoff
  \newpage
\fi

 \end{document}